\documentstyle[12pt]{article}
\begin{document}
\title{Precession of the Equinoxes and Calibration of Astronomical Epochs}
\author{B.G. Sidharth\\
International Institute for Applicable Mathematics \& Information Sciences\\
B.M. Birla Science Centre, Adarsh Nagar, Hyderabad - 500 063, India}
\date{}
\maketitle
\begin{abstract}
Astronomical observations were used as a marker for time and the
Calendar from ancient times. A more subtle calibration of epochs is
thrown up by an observation of the position of the solstices and
equinoxes, because these points shift in the sky with the years
resulting in the gradual shift of celestial longitudes $\lambda$.
Chronology based on such observations however needs to be backed up
by hard evidence. We match both to take us back to $10,000$ B.C.,
the epi-paleolithic period, and the beginning of civilization
itself.
\end{abstract}
\section{Introduction}
It is well known that due to the tilt (of $23 \frac{1}{2}^\circ$) of
the equator to the ecliptic, in other words the tilt of the axis of
the earth's spin to its plane of revolution, there is the phenomenon
of precession. This manifests itself in the Vernal and Autumnal
equinoctial points sliding in the reverse along the zodiacal belt
\cite{baker,payne}. The dynamical reason for this was explained by
Newton in the seventeenth century, while the period of revolution of
the equinoxes is roughly $25,800$ years. In star charts, this
manifests itself as the continuous change of longitude $\lambda$.\\
These effects can provide a marker for epochs in the following way:
In ancient times the equinoctial or solstice points were observed
along the solar and lunar zodiac. The lunar zodiac lies along the
solar zodiacal belt, but is made up of twenty seven parts identified
by the same number of lunar asterisms which have been used in
ancient India and subsequently in China and also much later in the
Arab world. (This corresponds to the $27 \frac{1}{3}$ days sidereal
period of the Moon. Each asterism or Nakshatra spans roughly
$13^\circ$). On the other hand it is relatively easy to pinpoint the
equinoctial and solsticial points, as these are characterized by,
respectively equal days and nights and longest or shortest days.
Thus historical observations of these phenomena can yield valuable
clues to the dates of observation. Final proof, however would have
to come from archaeo-astronomical evidence. With this in mind we
consider various astronomical references in ancient Indian
literature and then link them to actual archaeological evidence that
has been unearthed in recent times.
\section{Calendaric Astronomy and Astronomic Dating}
According to generally accepted ideas, civilization and science
began in Egypt and Sumeria in the third or fourth millennium B.C.
and spread in various directions \cite{stavrianos}. In the context
of the Indian subcontinent (that is, South Asia), it is believed
that Indo-Aryans, an Indo-European people, invaded the northwestern
parts of the country somewhere around 1500 B.C., overrunning in the
process the then existing Indus Valley or Harappan civilization
\cite{mazumdar}. The Indo-Aryans, so the theory goes, were a
semi-nomadic, hardy, rustic, and illiterate lot who could overcome
the civilized and settled Harappan inhabitants, destroying their
dwellings in the
process, because of their superior strength and equestrian skill.\\
This scenario is based on an interpretation of the earliest extant
Indo-European text, the R\d{g} Veda. The R\d{g} Vedic hymns are
supposed to be invocations to various tribal or naturalistic deities
for their aid in the Aryans' battles to conquer the original inhabitants.\\
Recently this view has been severely criticized, for various
reasons, and today is considered questionable and dubious. The
R\d{g} Veda and the related Vedic literature, on the contrary,
contain amazingly accurate and sophisticated calendric astronomy.
This fact points to not an illiterate, seminomadic tribal society,
but rather a well settled agrarian and meticulously scholarly
people. This is because, a calendar is required for agriculture
which in turn requires settlement. Once the astronomical content of
Vedic literature is recognized, several dates begin to tumble out
blatantly contradicting the prevailing picture of prehistory. It
will then be shown that very recent archaeological excavations
spanning a period of nearly $8500$ years, from about $10000$ B.C. to
about $1500$ B.C., can be meaningfully understood against this
background.
\section{The Calendric Astronomy of the Vedas}
The R\d{g} Veda repeatedly emphasizes \d{r}it\'{a} or cosmic order,
embodied in the periodicity of astronomical phenomena; the word for
seasons, \d{r}it\'{u}, derives from \d{r}it\'{a}. Both a solar and
lunar calendar were used, with an intercalation to reconcile the
two. This is true even today, in India, for the purpose of
festivals. Thus, the wheel of time has 12 parts and 360 spokes or
days or 720 pairs of day-nights, with a remainder of about 5 days
(RV 1.164). This is evidently a solar calendar. Interestingly, the
Egyptians had a 360-day year followed by a 5-day gap.\\
From time immemorial Vedic Indians have been using a luni-solar
calendar. The origin of this calendar is again to be found in the
R\d{g} Veda. Thus V\'{a}ru\d{n}a, (literally, "the
all-encompassing") knows the 12 moons (that is, the twelve, 29.5-day
months of a normal lunar year of 354 days). He also knows the moon
of "later birth" (that is, the thirteenth intercalated month, added
periodically to reconcile the lunar year of 354 days with the solar
year of about 365 days [RV 1.25]). The Sumerians also used such a
luni-solar calendar.\\
This gives an inkling of the problem at hand. A primitive society
cannot be expected to calculate the year or the lunar month or other
astronomical periodicities to any great degree of accuracy. The real
problem of the calendar is the following: the Moon is, to quote the
R\d{g} Veda, the maker of the month. Thus the interval between two
successive full moons, called the synodic month (or roughly, the
month), is 29.5306 days (more technically, mean solar days). But the
time taken by the Moon for one complete revolution about Earth,
called the sidereal month, is 27.3217 days. Finally the year that we
use, called the tropical year, that is, the year of the seasons, is
365.2422 mean solar days
\cite{illingworth}.\\
The problem now is to devise a convenient calendar that avoids the
fractional days and also the fact that twelve months do not exactly
fit the year.\\
I would like to stress at this point that any society meticulous in
its astronomical observation would choose the Moon quite naturally
as a calibrator for observation because the movement of the Moon in
the sky can be marked quite accurately against the background of the
stars, something that is not true for the Sun. So in the earliest
Vedic period, the path of the Moon was measured by twenty-seven
stars or star groups. These are the twenty-seven n\'{a}kshatras or
lunar asterisms. The Moon would spend one day in each n\'{a}kshatra.
But since the Moon takes 27.3 days for a complete circuit, at an
early point in time a twenty-eighth n\'{a}kshatra was also
considered. These n\'{a}kshatras were the daughters of the sky who
were wooed by the Moon. Some of these n\'{a}kshatras are explicitly
mentioned in the \d{R}g Veda, while the Yajur Vedas and the Atharva
Veda name them in detail. The \d{R}g Veda (10.85) says, "S\'{o}ma
[the Moon] in the midst of all these n\'{a}kshatras hath his place
...the Moon is that which shapes the years... S\'{o}ma [the Moon]
was he who wooed the maid [n\'{a}kshatra]..."\\
The n\'{a}kshatra system endured for a very long time (to this day,
in fact) and is seen clearly in the astronomical text Jyotisha
Ved\={a}\.{n}ga, (circa 1350 B.C.). The Chinese used the
twenty-eight n\'{a}kshatra zodiac called the shiu, an improvement on
the twenty-three n\'{a}kshatra scheme of about 850 B.C. of the
Yuehling. Some decades ago there was an unnecessary debate on the
origin of
the n\'{a}kshatra system \cite{macdonell}.\\
As can be clearly shown, its roots go back to the Vedas. An
important piece of evidence is the Weber manuscript, an antique
document discovered around 1890 at Kugiar near Yarkand in Sinkiang
Province. It describes the twenty-eight n\'{a}kshatras of the
Atharva Veda as expounded by an Indian scholar Pushkaras\'{a}rdi, in
the archaic Gupta script \cite{roy}. Recently it was widely reported
that an old star map on wooden blocks was salvaged from a pagoda
\cite{statesman}. It measures 25 centimeters by 21.2 centimeters and
has been dated at A.D. 1005, making it China's oldest star map. It
antedates by about onehundred years a previous star chart that was
discovered in 1971 in a tomb at Xuan Hua, in northern China. The new
find depicts the twenty-eight n\'{a}kshatras in a Sanskrit
incantation. (A recently discovered tomb painting from about 25
B.C., in Xian, appears to depict the twenty-eight n\'{a}kshatras
(hiu), while the earliest mention of their names might be traced
back to about 433 B.C.) The lunar n\'{a}kshatra system could also be
the symbolism behind the Arabic crescent moon
and star.\\
The Sun, however, cannot be ignored because the year is a barometer
of the seasons that regulate agriculture and all of life. Hence for
an agrarian community, the
need to reconcile the lunar months with the solar year.\\
A further remark is that in Vedic literature, the devas, or gods
(literally, "bright ones"), denote daytime, or the bright lunar
fortnight, or the northern course of the Sun, that is, from the
winter solstice to summer solstice. The asuras (or demons),
literally those prohibited from the celestial draft (of brightness),
denote the night, or the dark fortnight, or the southern half of the
Sun's annual course. The \.{S}at\'{a}patha Br\={a}hma\d{n}a
(2.1.4.9) declares that the gods are indeed the day
\cite{eggeling}.\\
Further, the so called Vedic rituals, as pointed out by the author,
were really observations (a view endorsed by the Late astronomer,
Prof. K.D. Abhyankar, as well.) A hint to this is given by a Rik (or
loosely part of a hymn), which declares that these Riks are written
in the high heavens, wherein are situated the bright ones. It goes
on to say, who will understand them who know not this, but we who
know it are assembled here. We will see dramaticf evidence
for this.\\
Finally, not only are Vedic astronomical computations incredibly
accurate, but they also show a way with members that borders on the
mystic and magical.\\
A very simple example is the concept of Gandharvas in the \d{R}g
Veda and Vedic literature in general, a concept that has been to
date grossly misunderstood. The Gandharvas are associated with the
Moon or S\'{o}ma. Indeed, they observe all forms (or phases) of the
Moon. According to the Aitareya Br\={a}hma\d{n}a (5.27), S\'{o}ma
(the Moon) lived among the Gandharvas, who returned the Moon in
exchange
for a woman (that is, a n\'{a}kshatra).\\
Their number is given variously as 27 and 6,333. All this is
perfectly and exactly meaningful if we realize that the Gandharvas
represent synodic months. First, 27 synodic months (from full moon
to full moon) approximately equal two years, whereas 6,333 synodic
months equal 512 years very accurately. This last relation in fact
gives the value of the synodic month as 29.5285 days and the year as
365.2422 days, revealing an incredible degree of accuracy.\\
Further, 6,333 synodic months equal 6,854 sidereal months. This
means that if a year began with the full moon in a particular
n\'{a}kshatra or lunar asterism, after 512 years, the year will
again begin with the full moon in the same lunar asterism.\\
There is a further twist; 512 is equal to $8^3$ and the well-known
G\={a}yat\={r}i meter of the \d{R}g Veda is an 8/3 meter. The \d{R}g
Veda declares that the G\={a}yat\={r}i meter has different
functions. For example in an 8-year period,
there are 3 intercalary months.\\
Another characteristic of the meter is in the above relation of
there being $8^3$, or 512, years in 6,333 synodic months. It can now
be seen why, in the \d{R}g Veda, the Gandharva is called
Vi\d{s}v\'{a}vasu or the universal V\'{a}su, the term V\'{a}su being
associated with the number eight. It is specifically stated that the
V\'{a}sus are associated with the G\={a}yat\={r}i meter.\\
Let us consider for another example the step-by-step build-up of
days, months, and years to the cycle or Mah\={a}yug\'{a} in Vedic
literature of 4,320,000 years. This Mah\={a}yug\'{a} or Megacycle is
half a cycle and encompasses the eclipse saros and the precessional
cycle as well. As I pointed out, there are 86,400 seconds in a day
and, furthermore, 4,320,000 equals $1^1 \times 2^2 \times 3^3 \times
4^4 \times 5^5$. All this is to elucidate my remark about the mystic
way with numbers that the Vedic composers had, in addition to their
amazing degree of accuracy in astronomical observations and insight
into astronomy itself. (It can be seen that the Babylonian eclipse
saros and the Greek Metonic cycle are not only included in the above
scheme, but far surpassed. In fact, the M\={a}\d{n}\d{d}\={u}kya
Upanishad specifies the 19/7 relationship of the Metonic cycle.)
\cite{gambirananda}. It is in this context that we can understand
the statement in the Jaimin\={i}ya Br\={a}hma\d{n}a, "Praj\={a}pati
[the lord of the devas or the bright halves, explained above]
defeated M\d{r}tyu [the lord of the asuras, the dark halves,
explained above] by numerical equivalence."\\
It is quite evident that the composers of Vedic literature were a
highly intelligent, knowledgeable, and sophisticated lot with a long
tradition of astronomy that itself implies observation and
settlement. Apart from their incredible calendric accuracy,
including knowledge of precession, such other advanced and modern
concepts as the heliocentric theory are already expected of the
seminomadic, illiterate invaders that contemporary theory supposes.
On this score alone, modern theories of Aryan invasion based on the
usual interpretation of the \d{R}g Veda and Vedic literature become
totally untenable. It is not surprising therefore that in recent
times emerging evidence has steadily eroded this theory
\cite{bgs}.\\
Once the astronomical and calendric character of Vedic literature is
recognized, any number of astronomical dates tumble out. All of them
show a continuous astronomical tradition beginning before 10,000
B.C. Many of these dates are couched in the typical allegorical
style of Vedic and other ancient Indian literature. But there are a
number of explicit dates also. A few are given below.\\
The Taittir\={i}ya Br\={a}hma\d{n}a (3.1.2) refers to
Aj\`{a}\'{e}kapada, the n\'{a}kshatra Purva Bh\'{a}pada, rising
exactly due east, a phenomenon that occurred around 10,000 B.C. when
this asterism was at the autumnal equinox.\\
Another explicit reference is from the Taittir\={i}ya Sa\d{m}hita
(6.5.3), which explicitly puts the asterism K\d{r}\'{i}ttik\={a}, or
the Pleides, at the winter solstice, an event that took place around
8500 B.C. While this was noticed by Dikshit in the past century, it
was dismissed as being an impossibly old date \cite{dikshit}.\\
The Aitareya Br\={a}hma\d{n}a, which is one of the earliest of the
Br\={a}hma\d{n}as or expository texts in the Vedas, explicitly
refers to the asterism P\'{u}narvasu (Castor and Pollux), presided
over by the deity \'{A}diti, also being exactly due east. This
happened around 6000 B.C., and was noticed by Tilak, but for
different reasons \cite{tilak}.\\
The \.{S}at\'{a}patha Br\={a}hma\d{n}a, oneof the latest
Br\={a}hma\d{n}as, refers to the asterism K\d{r}\'{i}ttikk\={a} (the
Pleides) rising due east, which immediately yields the date of about
2300 B.C. Thus an amazing continuity of astronomy from about 10,000
B.C. to about 2500 B.C. can be seen in Vedic literature.\\
All this is in blatant contradiction to contemporary theories that
the \d{R}g Veda was composed around 1500 B.C. by the invading
Indo-Aryans (who displaced the settled and civilized inhabitants of
the Harappan civilization). One could ask what is the evidence for
either the above date of the \d{R}g Veda or the invasion theory?
Surprising as it might seem, there is in fact practically no
evidence, archaeological or textual to support this claim
\cite{bgs2}. As will be seen, there is far more positive
archaeological, textual, and other evidence pointing to a very old
date for the Vedic civilization.
\section{Recent Geophysical and Other Hard Evidence}
At the time the history of antiquity was formulated, and until quite
recently, it was believed that about 10,000 years ago Earth was
under the grip of the last great Ice Age, which enforced a nomadic
lifestyle on people. Civilization in the modern sense of the word
began only after this Ice Age started thawing, which in turn made
agriculture and a settled lifestyle possible. This led to the great
river-valley civilizations in Egypt and Sumeria.\\
Recent geophysical studies of the stabilization of sea levels
indicate, however, that even around 10,000 B.C. several parts of
Earth had already warmed up to the extent of making agriculture
perfectly feasible \cite{schwartzberg}. In fact, in recent years it
has been realized that the epipaleolithic civilization of about
10,000 B.C. and earlier in and around Anatolia (Turkey) already
showed traces of agriculture (the growing of grain), domestication
of animals (such as goats and sheep), and permanent settlements in
round houses with arrangements for storage of food. This itself was
preceded by settlements with seasonal camps, dating from about
14,000 B.C. \cite{redman}.\\
The most dramatic archaeological finding of recent times has been
the relatively unknown excavations at Nevali Cori in Anatolia
\cite{hauptman}. This site dates back to around 7500 B.C. Current
excavations there reveal even older underlying structures, showing
an even older -- at least several centuries older -- civilization.
While there are other settlements not too far away that go back to
around 10,000 B.C., or earlier, the Nevali Cori civilization is
unique in that it represents an already developed civilization with
Megalithic elements and meticulous architecture and planning. The
inhabitants were also a very artistic lot and several beautiful
limestone sculptures have been found. In fact, this archaeological
site contradicts straightaway the theory that civilization began in
Egypt and Sumeria around 3000 B.C.\\
The very remarkable feature of Nevali Cori is that in civilizational
terms it is an isolated oasis within the framework of present-day
knowledge. It does not relate to any civilization or culture of its
period. There is a gap of some five thousand years before we come to
a similar civilization. With one exception.\\
Its echoes can be found again in the fairly recent excavations at
Mehrgarh in the Baluchistan area of the Indian subcontinent
\cite{jarrige}. The Mehrgarh civilization dates back to between 7000
B.C. and 6000 B.C. and parallels Nevali Cori in terms of economy,
agriculture, domestication of animals, and the planning and layout
of large settlements (Professor Hauptmann, the excavator of Nevali
Cori, subscribes to this view).\\
As noted Professor Hauptman and colleagues had unearthed at Nevali
Cori near Sangli Urfa structures dating back to about 8000 B.C.
However the earliest astronomical date given based on the rising
exactly at the east point (vernal equinox) of the asterism
Aj\`{a}\'{e}kapad (Uttar Bhadrapada) was, as pointed out 10,000 B.C.
A few years after Hauptman's work, Klaus Schmidt and coworkers
discovered other very similar megalithic structures at Gobekli Tepe
barely a few kilometers from there which, as predicted go back to,
indeed 10,000 B.C. The pillars and motifs at Gobekli Tepe are
clearly in the same mould as those at Nevali Cori--the latter in
fact show a continuation. Once again these subsequently discovered
pillars and motifs have astronomical symbolism which can be
interpreted in terms of the \d{R}g Vedic symbols and concepts.
Firstly, in an enclosure ("D") there are twelve obelisks or pillars,
one for each month!  These pillars each show the figure of a fox or
wolf. The fox or Vrika in the \d{R}g Veda is the Moon, this being
another meaning of the same word. The \d{R}g Veda mentions that the
fox (or she wolf) "having seen me once, slouches away through the
houses." The same line also means, the Moon, the maker of the
months, moves through the houses. This is clearly Luni-Solar
astronomy. Each of the twelve foxes can be clearly seen to represent
a month! Similarly there is the symbol of the thin crescent Moon
being eclipsed by the Sun with a symbol resembling "H" above. This
again is a reference to a total solar eclipse in the constellation
of Gemini the twins.
There are other familiar astronomical symbols like the Scorpion.\\
Another striking motif on the Gobekli Tepe pillars is that of a bird
carrying the Sun representing Vishnu, the Sun riding atop the bird
Garuda. There is also a phallic motif with a man. This represents
the incestuous relationship which Prajapati (Orion) had with his
daughter, and has also been elaborated upon in the Aitreya Brahmana.
Because of this sin, Prajapati was shot with an arrow, represented
by the three waist stars of Orion. His head also was cut off and
replaced with the head of an animal, represented by the star
Betelguese or Arudra. All this has been elaborated in the Vedic
literature.\\
Perhaps the definitive fact associated with Gobekli Tepe is the
following. There are the hymns in the Rig Veda of Sunashepa. He was
to be sacrificed, and tied down with three pegs (the three stars of
the Orion belt). One kept him on Earth, one was in the heavens and
one was in the middle. Thus was Sunashepa tied down the hymns being
inexplicable. The remarkable fact is that at the latitude of Gobekli
Tepe, around 10,000 B.C., not only was the winter solstice near the
Orion -- Taurus system, but also one of the three stars was below
the horizon, one near the horizon and the third above the horizon.
This is a dramatic confirmation of date and place with pure
astronomical positions. Incidentally Taurus the bull (Vrishabha)
itself is a motif in the pillars there. Incidentally this theme of
the three belt stars of Orion recur in different forms in ancient
Indian literature, as pointed out in detail, elsewhere -- as the
legend of the King Trishanku trying to enter the heavens, as the
hero of Mahabharata, Yudhistara entering heaven
with the dog, Sirius and so forth.\\
The buildings at Mehrgarh were constructed of mud bricks. Several
rooms were used for habitation and storage of food. Stone tools were
used for harvesting cereal grasses. Ornaments were also being made.
Social differentiation was also evident. By 5000 B.C., mud-brick
storage houses were built and public architecture began to appear,
as did handicrafts, including handmade pottery. Copper metallurgy
was also present. Soon barley and wheat were also being extensively
used. It is believed that between 5000 and 4000 B.C. the Mehrgarh
civilization expanded. Many mud-brick storage buildings were built
and now pottery was made using potter's wheels. There was a
continuation in the making of ornaments and metallurgical
activities. A few contemporary sites in Baluchistan and the
Northwest Frontier province also appeared in this period. An article
of pottery here from the period 4000 B.C. to 3500 B.C. shows the
svastika symbol in a circle.\\
From about 3500 B.C. the Mehrgarh civilizational influence began to
spread, particularly in pottery technology and styles, linking not
only Central Asia (southern Turkmenistan), including the southern
parts of Afghanistan and eastern Iran, but also the Harappan
civilization. Stamp seals began to appear. In fact, the various
interactions in the region are believed to have set the stage for
the emergence of the Harappan civilization that, until quite
recently, was supposed to have had a sudden and independent origin
in full bloom!\\
The Harappan civilization is much too well known to be elaborated
upon here \cite{schwartzberg}. It existed from about 2500 B.C. to
about 1800 B.C. based on carbon dates (or from 3100 B.C. to 1900
B.C. with MASCA correction). Nearly one thousand Harappan sites have
been unearthed spanning a vast area of 1.5 million square
kilometers, from the borders of Iran, to Turkmenistan (Altyn Depe),
and from northern Afghanistan (Shortughai in the Bactrian Plain),
through Punjab and Gujarat right up to Delhi and the Godavari
Valley. It is now believed that this civilization developed from the
early Indus Valley groups, especially Kotdiji, which itself was
influenced by Mehrgarh. This vast civilization displays a remarkable
uniformity mirroring the settlement patterns of the towns of
Mohenjodaro and Harappa.\\
It is now clear that fire worship was prevalent in this civilization
\cite{srrao}. Brick altars for fire worship were built in many
houses, and in Kalibangan a row of seven fire altars was found,
while one of the Harappan seals depicts worship of the fire god with
seven attendants. A few seals showing a horned deity in yogic
posture amidst animals have been interpreted, probably correctly, as
depicting the god \'{S}iva in his aspect of Pa\.{s}\'{u}p\'{a}ti,
lord of animals.\\
There does not appear to have been a major capital or center for
this vast civilization, quite unlike other ancient civilizations.
This suggests what has been described as a complex chiefdom, or a
series of chiefdoms strung together, rather than a unified state.\\
The very developed public architecture of the Harappan civilization
is well known, as are the thousands of seals. The decipherment of
the script of these seals has been a problem. It was believed to
have been a form of an early Dravidian language, but it now appears
that the language of the seals was what the Indian archaeologist Rao
calls an old Indo-Aryan, more plainly, a form of Sanskrit
\cite{kak}.\\
The Harappans had much contact with several places in Iran,
Tajikistan, Bahrain, and elsewhere, as is evidenced by the Harappan
pottery, weights, and seals found in those places.\\
It was believed that the Harappan civilization suddenly fell or
disappeared. It now appears that many Harappan sites were affected
by floods and by rivers changing their courses, and this led to the
abandonment of settlements, in spite of the many low, protective
walls built against such an eventuality. Though Harappan
civilization declined, it continued in a decadent form in Gujarat
and other parts of India. This phase is being labeled as the late
Harappan period. Such late sites are found as far south as the
present-day Maharashtra and as far east as Delhi. They extend up to
the late second millennium B.C., and at these sites the painted
grayware pottery of the early Iron Age is also found, indicating a
direct cultural link between the Bronze and Iron Ages in the Indian
subcontinent. This belies an earlier theorized Dark Age. The
once-popular theory that the inhabitants of the Harappan
civilization fled due to invasion is now called into question.\\
Around 2000 B.C. the cemeteries and tombs of the Mehrgarh region
show some new features that are also found in the late Harappan
civilization. These new traits are similar to those of Tepe Hissar
III near Iran and Namazga V of southern Turkmenistan, and also to
regions in Afghanistan like Dashly III. It has now come to light
that around 2000 B.C. a vast region extending from the Gurgan Plains
near the Caspian Sea (for example, Tepe Hissar) through southern
Turkmenistan (Namazga, Altyn Depe), the Murghab Delta (ancient
Margiana, Togolok, and so on), and ancient Bactria, including north
Afghanistan (Dashly) right up to Mehrgarh in Baluchistan show
uniform cultural traits. Parpola calls this the Bronze Age of
"Greater Iran" or Namazga V culture \cite{parpola}. It is
characterized by distinctive monumental architecture, cult objects,
iconographic motifs, bronze swords, and so on. Links with the
Harappan civilization are also evident.\\
Further, the large quantity of weapons and evidence of chariots
indicate the presence of a military elite -- an Aryan military
elite, as is generally agreed. At Dashly III a fire temple has come
to light (circa 2000 B.C.) that has three concentric circular walls,
and there is evidence of another such temple as well. At Togolok 21
another fire temple has been unearthed (circa 1800 B.C.). Apart from
the fire altars, at the Togolok temple the ancient Vedic S\'{o}ma
ritual was also practiced; the oblations were put in a row of
vessels placed on special brick platforms. Further, at this site
thirty miniature stone columns were discovered, as have been found
at many other sites of Greater Iran. These are supposed to provide
the link between the phallic cults of West Asia and the Hindu
li\.{n}ga cult also in evidence in the Harappan civilization.\\
Apart from all these recent and comparatively recent developments,
the existence of Aryan tribes in and around Anatolia -- the Mittani,
the Hittites, the Kassites, and so on, who lived around the middle
of the second millennium B.C. -- has been known for a long time.
Thus the Vedic deities, \'{I}ndra, Mitra, V\'{a}runa, and
N\'{a}satya are invoked in the Boghuz Koi (near Istanbul) oath
inscriptions between the Mittani king Sati Vaja and the Hittite king
Supliluliuma \cite{piggott}. In addition, the prevalence of other
Vedic deities like the M\={a}rutas, S\={u}rya, and also the
horse-racing terms in Kikkuli's manuscript all show a distinct Vedic
and Sanskritic (Indo-Aryan) influence. There have been a number of
theories about these civilizationsof antiquity (but not of the newly
discovered Nevali Cori civilization).\\
With regard to the Nevali Cori civilization, as I have pointed out,
in addition to a remarkable coincidence of dates, from several
points of view there is very good reason to identify the Nevali Cori
civilization with the Vedic \cite{bgs3}.\\
With regard to the Harappan civilization, there has been a theory
popular among European scholars for several decades, namely, the
Harappans were originally some sort of Dravidians and that around
1500 B.C., invading Indo-Aryans, who were also the composers of the
\d{R}g Veda, overran the Dravidian Harappan civilization and the
Harappans were driven to the southern parts of India. For some of
the reasons discussed in detail earlier (Cf.ref.\cite{statesman} for example), this theory
has been called into question severely.\\
\section{Astronomy and Related Considerations}
With this general scenario we can now discern a number of links,
which form a suggestive mosaic:\\
1. The svastika symbol in Mehrgarh pottery, which appears some what
later on some Indus seals, is an auspicious symbol in Indian epic
literature. In addition, the Mehrgarh civilization dates back to the
seventh millennium B.C. and has aspects of similarity with the
Nevali Cori civilization.\\
2. There seems to be a connection between the fire altars in
Turkmenistan (Togolok) and Afghanistan (Dashly) and the Harappan
civilization, particularly Kalibangan, where there are seven fire
altars, and also with the Harappan seal showing worship at a fire
altar with seven accompanying deities.\\
The concept of the seven fires is purely astronomical and originates
in the \d{R}g Veda. It is connected with the myth of the stars of
the Pleiades or K\d{r}\'{i}ttik\={a} and the seven stars of the
Great Bear or Sapta \d{R}\'{i}shi (the Seven Sages), and its
exemplified in the Mah\={a}bh\={a}rata. To understand this myth one
has to notice that of the seven visible stars in the Great Bear,
only one has a companion. There are six easily visible stars in the
Pleiades. According to the myth, the fire god (that is, the Sun) was
enamoured of the seven wives of the seven sages. Another maiden
(Sv\={a}h\={a}) was enamoured of the fire god. So Sv\={a}h\={a}
successfully took the form of the wives of the sages, that is, the
Great Bear, and cohabited with the Sun. Except that the wife of one
of the seven sages was so chaste that Sv\={a}h\={a}just could not
take her form. The six wives were promptly banished. In any case the
resulting semen was put at one place --this is the Pleiades, which
according to the dates proposed above was at the winter solstice in
early Vedic times. To conclude the myth, the result of the
cohabitation was the Hindu deity K\={a}rttikeya, who was split
asunder, the second portion being the lunar asterism
V\'{i}\d{s}\={a}kha, that is, one who has been split. (As pointed
out elsewhere, the lunar asterism V\'{i}\d{s}\={a}kha is 180 degrees
away and would have been at the summer solstice in the Vedic epoch.)
K\={a}rttikeya, born of the Pleiades, rides a peacock, probably a
symbol for the several stars that constitute this lunar asterism. He
is nurtured by the seven mothers. To make the identification of the
Indus seal script complete, Rao's decipherment shows that the fire
altar on the seals was called {\it gahpahppat}, which clearly
denotes g\={a}rhapatya, the Vedic fire altar to be kept at home.\\
What would be the significance of the Sun in the form of a fire god,
the Pleiades, and the seven sages and seven mothers for the Harappan
civilization fire altar? That at the time of the Harappan
civilization the Pleiades was at the vernal equinox, as in fact is
explicitly mentioned by the latest of the Br\={a}hma\d{n}as, the
\.{S}at\'{a}patha Br\={a}hma\d{n}a.\\
It is quite remarkable that the peacock and stars are prominent
motifs on the pottery found in the late Harappan cemetery $H$ which
pottery already shows new stylistic influences \cite{bgs5}. Among
the objects found there was an urn or pot with seven seedlike
objects in it that can be directly compared with the above myth. So
while the styles may have changed, the concepts seem to show a
continuity.\\
The seven sages and seven mothers and seven seeds have an exact
parallel in the Zoroastrian Haptaoring (in Sanskrit saptali\.{n}ga
or the seven phallic objects), and the seven sages of Sumerian lore.
There is an equal parallel with the seven deities of the Hittites.
In fact, the parallel is made closer if we observe that in the
Hittite custom a goat approaches the deities. In the Pleiades myth
from the Mah\={a}bh\={a}rata related above, a goat is also present.
The link with the Hittites is of course quite natural. Apart from
the many other known parallels, recently Parpola has pointed out
another curious similarity between a chariot drawn by a horse and a
mule in Hittite custom and the identical ancient Hindu vipatha
chariot. Also, a few Hittite invocations are practically in
Sanskrit. For example, the Hittite "Istanue Ishami," compares
exactly with the Sanskrit, "Sth\={}nu Is'ami." \cite{dehiya}.\\
Similarly, from very early on a deity seated in yogic pose amid
animals depicted on Harappan seals has been identified with the
Hindu \'{S}iva. This again throws up a contradiction if it is
supposed that the Harappans were Dravidians or non-Aryans, because
\'{S}iva in Hindu tradition is a br\={a}hmin, very much an "Aryan"
concept. (Curiously enough, theplace where the Harappan civilization
was discovered in the past century by the British was called
Brahminabad or Brahmin City. Could this be more than mere
symbolism?) A final remark in this context is that the Harappans
used a year of six seasons, as in the \d{R}g Veda, and further used
a sixty-year (Jovian) cycle, which has been an ancient Hindu
tradition.\\
3. The goddess associated with a tiger in a Kalibangan cylinder seal
compares closely with a cylinder seal from Shahdad (Iran), also
depicting a goddess. This compares with seals showing a goddess on a
tiger (or lion) from Bactria \cite{parpola}. All this is easily
understandable in terms of Durg\={a} of Hindu mythology. This also
shows links with Sumeria, because until comparatively recently,
cylinder seals were exclusively associated with that region.\\
4. The eagle-headed deity from Baluchistan of the second millennium
B.C. compares with the motifs of similar deities fighting serpents
found in Greater Iran, with the variation of a human head and a bird
body (both considered to be the same), again fighting serpents
\cite{parpola}.\\
5. The Harappan seal found at Altyn Depe shows a link between
central Asia and Harappan civilization \cite{bongard}. In fact,
Prof. V. Masson on this basis suggested that the south central Asian
people were, like the Harappans, Dravidian.\\
6. The pottery and terracotta figurines from southern Turkmenistan
closely resemble similar objects from the Harappan civilization
\cite{bongard}.\\
7. The Vedic deities and Indo-Aryan names of Anatolia of the second
millennium B.C. evidently show a close linkage.\\
8. The inhabitants of Greater Iran were Aryans while the language
and culture of the Harappan civilization is also of the same origin.
Furthermore, the racial types of southern India are no different
from the Caucasian and Mediterranean types.\\
9. The burial material in southern Tajikistan of the second
millennium is close to those in Indo-Aryan burials \cite{bongard}.\\
10. All this can be tied up with the distinct Vedic influence at
Nevali Cori in the form of a limestone sculpture of a head of a
Vedic priest, going back to the eighth millennium B.C.\\
The picture that emerges from the above mosaic of evidence and dates
is that in the eighth millennium B.C. or earlier a Vedic culture
with \d{R}g Vedic Sanskrit as its language was already apparent in
Anatolia, and that it gradually diffused toward both east and west.
For a few thousand years --may be about five thousand years--the
language, religion, and culture gradually evolved and changed. This
picture is in harmony with very recent findings based on blood
groups and languages--in fact, from Europe through the Mediterranean
regions to India we have the ethnic caucasoid group \cite{cavali}.\\
A model having some resemblances to the above was proposed rather
recently by Renfrew \cite{renfrew}. But Renfrew's model conflicts
with that of French scholar Dum\'{e}zil \cite{facets} who had argued
that the several similarities in the myths of diverse and very
widespread Indo-European cultures could not be merely accidental or
due to routine diffusion and contact, as is implied by Renfrew.
Indeed this would be far-fetched. Dum\'{e}zil, on the lines of Max
M\"{u}ller, went on to trace a common ancestry for the Indo-European
people among the Kurgan Russians, circa 3000 B.C. Renfrew and
Dum\'{e}zil completed their work before some of the excavations and
other evidence touched on in this discussion. The scheme proposed
here, in a sense, reconciles these two views to some extent.\\
In the earlier models, however, the language of the Anatolian region
in the eighth millennium was a hypothetical proto-Indo-European and
not \d{R}g Vedic Sanskrit. Apart from the doubts cast on
glotto-chronology, it is possible that the rate of change of
languages was slower the farther back in time we go. Such a very
slow and relatively peaceful diffusion is in fact mirrored by the
gradual development of the Mehrgarh civilization. It is also
vindicated by the Harappan model of a vast civilization comprising a
large number of coexisting chiefdoms.\\
This brings us to around the second millennium B.C. where again we
see a vast number of connections. This is also the period of the
latest Br\={a}hma\d{n}as and the Upanishads and the earlier epic
period of Hindu literature. Indeed the literature portrays a picture
not so much of a centralized empire, but that of a number of loosely
interlinked smaller kingdoms, in relative harmony, on the lines of
the Harappan settlements. In fact, the names of a number of Hindu
epic dynasty families, for example, the K\'{u}rus, can be traced to
Greater Iran. Similarly tribe or community names from Hindu epics
like the V\d{r}ikas can be traced right up to the Gurgan Plain
\cite{parpola}. The far-extending lion-or tiger-borne mother goddess
motifs, the eagle and serpent motifs, and the svastika motifs all
support this view and can be understood against this background.
Further, the "Old Indo-Aryan" language of the Harappans, in the
context of the above scenario, suggests that the language, rather
than being a precursor of Vedic Sanskrit, was a corruption of the
latter due to passage of time and distance; the Indus Valley may not
have been the epicenter of the then prevalent Vedic civilization,
which was probably centered in adjacent Greater Iran.\\
There is in fact astronomical evidence for the above assertions. Two
ancient Hindu astronomical texts, the Jyotisha Ved\={a}\.{n}ga and
the Pit\={a}maha Siddh\={a}nta (attributed to the hero scholar
Bh\={i}shma of the Mah\={a}bh\={a}rata), both dating to around the
middle of the second millennium, are set in a latitude of about 35
degrees north, as can be deduced from the given relative length of
the longest day of the year. This falls right inside Greater Iran. A
few scholars have taken the unfounded and untenable view that this
merely represents a trace of Babylonian astronomy. Others would be
reluctant, again without much of a basis, to transplant the epic
Hindu setting to Greater Iran. But the fact is that in the
Mah\={a}bh\={a}rata) itself, places such as Gandh\={a}ra, Kamboja
(Afghanistan), and Sindh, and peoples such as the Yavanas, Valhikas,
and so on, are featured. According to the longitude given by
Var\={a}hamihira, circa 500 B.C. Yavana (probably Ionia) would
correspond to Alexandria while Valhika has been identified with
Balkhash \cite{thibaut}.\\
Large-scale and possible violent migrations or invasions seem to be
a feature more evident from the second millennium B.C. According to
Soviet archaeologists, it appears that only from the middle of the
second millennium B.C. did active migrations of the steppe tribes of
central Asia take place, and these tribes penetrated the erstwhile
farming and settled cultures \cite{bongard}. This situation is also
mirrored in west Asia around the same time.
\section{Concluding Observations}
The very ancient date of around 10,000 B.C. proposed for the \d{ R}g
Veda or Vedic culture now appears plausible in view of the
epi-Paleolithic agricultural or proto-agricultural civilizations
dating back to a similar or even earlier period. If, as European
scholars have supposed, the \d{R}g Vedic descriptions of invasions
of the circular forts of the d\={a}sas or aborigines are to be taken
at all literally, couldn't these forts be identified with the
circular dwellings of the Neolithic people in and around the
Anatolian region?\\
In fact, the word Aryan of the \d{ R}g Veda is derived from the
Sanskrit root meaning "to plow." What all this is about is lucidly
expounded in several Pur\={a}\d{n}as (for example, the Vish\d{n}u
Pur\={a}\d{n}a) in their characteristic allegorical style. In a
nutshell: From the thigh of King Vena, all the evil came out in the
form of a black dwarf (that is, an aboriginal pygmy). From the
king's right hand, came out a beautiful shining prince,
P\d{r}ith\'{u}, who, because of a famine, pursued Earth intending to
slay it, as it would not yield its fruits. Earth finally relented.
"Before his time there was no cultivation, no pasture, no
agriculture, no highways for merchants, all these things originated
[then]... Where the ground was made level, the King induced his
subjects to take up their abode....Then proceeded all kinds of corn
and vegetables upon which people now subsist...." \cite{wilkins}.
P\d{r}th\'{u} 's son was Manu of Hindu Mythology, the progenitor of
humankind. Thus the "Arya" of the \d{R}g Veda would represent,
rather than an ethnic type, the very first agricultural people
whence civilization itself began, sometime prior to about 10,000
B.C. \cite{nevalicori}.\\
In this connection, it is interesting to note that the term Aryan
with its modern connotation is of rather recent coinage. Once the
similarities between Sanskrit, Greek, Latin, German, and the Celtic
languages were discovered by Sir William Jones, the term
"Indo-Germanic" was coined by Bopp, the nineteenth-century German
comparive philologist. This was later rechristened "Aryan", by Max
M\"{u}ller, from the \d{R}g Vedic --supposedly racial--"Arya"
\cite{mackenzie}.\\
But the \d{R}g Veda and Vedic literature show a level of astronomy
and science far ahead of that seen from about 2000 B.C. onward.
Historians have, by andlarge, adopted a linear model of progress
that has not always been true. There have been periods of regression
in human civilization and human knowledge. It appears that very
advanced but camouflaged astronomy in the \d{R}g Veda slowly
decayed, as its meaning was lost. In fact, even the earliest of
Br\={a}hma\d{n}as like the Aitareya Br\={a}hma\d{n}a (circa 6000
B.C.) already begins to speculate on the possible meanings of the
\d{R}g Veda.\\
But there is a curious linguistic feature that might just symbolize
this bifurcation. The Judeo-Christian traditions are from Abraham
(of the second millennium B.C.), a word that could be interpreted as
being derived from a-brahman, with brahminism being identical to the
\d{R}g Vedic tradition.\\
As noted long ago by the author, this date and tradition exactly
matches the description given by Plato in Tinaeus about the lost
civilization, which he called Atlantis.


\begin{thebibliography} {99}
\bibitem {baker} Philippe de la Cotardiere (1987). \emph{Larousse Astronomy}
(Ed.Mark R. Morris) (Hamlyn, London), 1987.
\bibitem {payne} C. Payne-Gaposchkin (1954). \emph{Introduction to
Astronomy} (Methusen \& Co. Ltd., London), 1954.
\bibitem {stavrianos} Mallory, J.P. (1991). \emph{In Search of the Indo Europeans} (Thames \& Hudson,
London), 1991.
\bibitem {mazumdar} Mazumdar, R.C. (1991). \emph{Ancient India}
(Motilal Banarsidass, Delhi), 1991, p.28ff.
\bibitem {illingworth} Illingworth, V. (1981). \emph{A Dictionary of
Astronomy} (Pan Books, London), 1981.
\bibitem {macdonell} Macdonell, A.A., and Keith, A.B. (1982). \emph{Vedic Index
of Names and Subjects} (Motilal Banarsidass, Delhi), 1982, p.409ff.
\bibitem {roy} Roy, S.B. (1976). \emph{Ancient India} (Institute of Chronology, New Delhi), 1976, p.66.
\bibitem {statesman} Sidharth, B.G. (1999). \emph{The Celestial Key to the
Vedas} (Inner Traditions, New York).
\bibitem {eggeling} Eggeling, J. (1978). \emph{The Satapatha
Br\={a}hma\d{n}a} (Motilal Banarsidass, New Delhi), 1978.
\bibitem {haug} Haug, M. (1977). trans. \emph{Aitereya Br\={a}hma\d{n}am of the \d{R}g
Veda} Vol.2 (Bharatiya Publishing House, Varanasi), 1977, p.173ff,
p.212ff.
\bibitem {gambirananda} Swami Gambirananda. (1978). \emph{Eight
Upanishads} (Advaita Ashrama, Calcutta) 1978.
\bibitem {bgs} Sidharth, B.G. (1978). \emph{Glimpses of the Amazing
Astronomy of the \d{R}g Veda}; \emph{The Unmythical Puranas: A Study
in Reverse Symbolism} and \emph{Did Indians Pioneer Astronomy?} in
\emph{Indological Taurinensia} 6, 1978.
\bibitem {dikshit} Dikshit, S.B. (1969). \emph{Bharatiya Jyotisha Shastra}
(Govt. of India Press, Calcutta), 1969.
\bibitem {tilak} Tilak. B.G. (1955). \emph{Orion} (Tilak Brothers, Pune), 1955,p.215ff.
\bibitem {bgs2} Sidharth, B.G. (1990). \emph{Ancient Indian
Cosmology}, \emph{History of Indian Science and Technology}, Vol.2
(Sandeep Prakasan, New Delhi) 1990.
\bibitem {schwartzberg} Schwartzberg, J.E. (1978). \emph{A
Historical Atlas of South Asia} (The University of Chicago Press,
Chicago) 1978.
\bibitem {redman} Redman, C.L. (1978). \emph{The Rise of
Civilization} (W.H. Freeman, San Francisco) 1978.
\bibitem {hauptman} Hauptmann, H. (1991-92). \emph{Nevali Cori},
\emph{N\"{u}rnberger Bl\"{a}tter Zur Archaologie} (1991-92): 15ff.
\bibitem {jarrige} Jarrige, J.F. and Meadow, R.H. (1980). \emph{The
Antecedents of Civilization in the Indus Valley} \emph{Scientific
American} 243, 1980, 102ff.
\bibitem {srrao} Rao, S.R. (1991). \emph{Dawn and Devolution of the
Indus Civilization} (Aditya Prakashan, New Delhi) 1991, pp.272-281.
\bibitem {kak} Kak, S.C. (1992). \emph{The Indus Tradition and
Indo-Aryans}, \emph{Mankind Quarterly} 32, No.3, 1992, p.43; Hajra,
S. \emph{On the Decipherment of the Inscriptions of the Seals of
Harappa and Mohenjadaro} (Subarnarekha, Calcutta) 1974.
\bibitem {parpola} Parpola. (1988). \emph{The Coming of the Aryans to Iran
and India} (Studia Orientalia, Helsinki), No.64, 1988,pp.192-305.
\bibitem {piggott} Piggott, S. (1961). \emph{Prehistoric India}
(Penguin, Middlesex) 1961; Kak \emph{On the Chronology of Ancient
India}.
\bibitem {bgs3} Sidharth, B.G. (1982). \emph{A Lost Anatolian Civilization -
Is it Vedic?} (B.M. Birla Science Centre Research Communication,
Hyderabad), 1982.
\bibitem {srrao2} Rao, S.R. \emph{Excavation of Submerged Ports -
Dwaraka, A Case Study} Also Pusalker, A.D. \emph{Studies in Epics
and Puranas of India} (Bharatiya Vidya Bhavan, Bombay), 1955,
p.75ff.
\bibitem {roymaha} Roy, P.C. (1956). \emph{The Mahabharata} (trans.) (Oriental Publishing Co., Calcutta), 1956.
\bibitem {shaffer} Shaffer, J.G. (1984). \emph{The Indo-Aryan
Invasiions: Cultural Myth and Archaeological Reality} in \emph{The
People of South Asia} (Ed. J.R. Lukacs), (Plenum, New York) 1984.
\bibitem {bgs5} Sidharth, B.G. (1993). \emph{The Antiquity of the Rg
Veda} (B.M. Birla Science Centre Research Communication, Hyderabad),
1993.
\bibitem {dehiya} Dehiya, B.S. (1988). \emph{Aryan Tribes in West
Asia} , \emph{Vishveshvaranand Indological Journal} 36, 1988, 219.
\bibitem {bongard} Bongard-Levin, D. and Viogasin, A. (1984).
\emph{The Image of India} (Progress Publishers, Moscow) 1984, p.190.
\bibitem {cavali} Cavali-Sforza, L.L. (1991). \emph{Genes, Peoples
and Languages} in Vigyan \emph{Scientific American, Indian Edition}
December 1991, p.70ff.
\bibitem {renfrew} Renfrew, C. (1987).  \emph{Archaeology and
Language: The Puzzle Indo-European Origins} (Jonathan Cape, London),
1987.
\bibitem {facets} (1992). \emph{FACETS} (Em,bassy of France in India) No.1,
1992, 24ff.
\bibitem {thibaut} Thibaut, G. and Dvivedi, S. (1930). trans.
\emph{Panchasiddhantika of Varaha Mihira} (Motilal Banarsidass,
Lahore) 1930, p.19.
\bibitem {wilkins} Wilkins, W.J. (1988). \emph{Hindu Mythology}
(Rupa, Calcutta) 1988, p.15ff.
\bibitem {nevalicori} A historical find near \emph{Nevali Cori} has
just been reported by Dr. Gillian Vogelsang-Eastwood of the National
Museum of Ethnology in Leiden, Prof. Robert Braidwood of the
University of Chicago, and Prof. Frank Hole of Yale University.
Thisis a small piece of cloth found at Cayonu, dating back to 8000
B.C. (According to current ideas cloth weaving would go back to
around 3500 B.C.) This dramatic find fits in very harmoniously with
the date, culture and location of the \d{R}g Vedic civilization, as
proposed in this paper.
\bibitem {mackenzie} Mackenzie, D. (1996). \emph{Myths and Legends}
(Bracken Books, London) 1986, p.19.
\bibitem {reincourt} Reincourt, A. de. (1982). \emph{The Eye of
Shiva} (Souvenir Press, Condor) 1982.
\end{thebibliography}
\end{document}